\documentclass{elsart}

\usepackage{natbib}
\usepackage{amssymb}
\usepackage{graphicx}
\journal{Geomorphology}

\begin{document}

\begin{frontmatter}

\title{Profile measurement and simulation of a transverse dune field in the Len\c{c}\'ois Maranhenses}

\author{E. J. R. Parteli$^{1}$,}{\author{V. Schw\"ammle$^{1,2}$,}} \author{H. J. Herrmann$^{1,3}$,} {\author{L. H. U. Monteiro$^{4}$}}\and\author{L. P. Maia$^{4,5}$}
\begin{center}
\address{(1) Institut f\"ur Computerphysik, ICP, Pfaffenwaldring 27, 70569, Stuttgart, Germany.
	 (2) Instituto de F\'{\i}sica Te\'orica, Universidade Federal Fluminense, 24210, Niter\'oi, RJ, Brazil.
	 (3) Departamento de F\'{\i}sica, Universidade Federal do Cear\'a, 60455-970, Fortaleza, CE, Brazil.
	 (4) LABOMAR, Av. Aboli\c{c}\~ao, 3207, Meireles, 60165-081, Fortaleza, CE, Brazil.
	 (5) Departamento de Geologia, Universidade Federal do Cear\'a, 60455-970, Fortaleza, CE, Brazil.
	 }
\end{center}

\begin{abstract}
In this work, we report measurements of the height profile of transverse dunes in the coastal dune field known as ``Len\c{c}\'ois Maranhenses'', northeastern Brazil. Our measurements show that transverse dunes with approximately the same height present a variable brink position relative to the crest, in contrast to the case of barchan dunes. Based on our field data, we present a relation for the dune spacing as a function of the crest-brink distances of transverse dunes. Furthermore, we compare the measurements with simulations of transverse dunes obtained from a two-dimensional dune model, where a phenomenological definition is introduced for the length of the separation streamlines on the lee side of closely spaced transverse dunes. We find that our model reproduces transverse dune fields with similar inter-dune distances and dune aspect ratios as measured in the field.

\end{abstract}
\end{frontmatter}

%***************** SECTION 1: INTRODUCTION ********************************

\section{Introduction}

Dune morphology has been investigated over the last fifty years by scientists from different research areas like geologists, geographers, physicists and even mathematicians, since the pioneering work of \cite{Bagnold_1941}. Although a theoretical explanation for the wide variety of observed dune forms is still lacking, it is well known that the behaviour of the wind as well as sand availability are fundamental parameters for determining the final dune shape. On bedrocks and surfaces with not too much sand and under uni-directional winds, barchan dunes move separated from each other, without changing their shape and with a velocity inversely proportional to their height. These dunes have a windward side and a steep slip face where avalanches take place maintaining dune stability and motion. As the amount of available sand increases, ``barchanoids'' are found, which look like oscillating chains of barchans joined at their horns, aligned perpendicularly to the wind direction. For higher sand supply, transverse dunes are observed. Transverse dunes cover 40$\%$ of all terrestrial sand seas, being common in the Northern Hemisphere, particularly in China, and dominating the dune fields on Mars. Normally transverse dunes have more irregular patterns and even hierarchies of smaller dunes can be found on them. Transverse dunes have been studied in the field many times \citep{Wilson_1972,Lancaster_1982,Lancaster_1983,Mulligan_1988,Burkinshaw_and_Rust_1993,Wiggs_2001}, but still most of these works are restricted to desert dunes.

Dunes appear very commonly along coasts \citep{Hesp_et_al_1989,Nordstrom_et_al_1990,Hesp_2002,Kleinhans_et_al_2002,Goncalves_et_al_2003,Barbosa_and_Dominguez_2004}. The sand that constitutes these dunes comes from the sea being thereafter deposited on the beach. Once the grains are exposed to the air, they dry and some can be carried by the wind, initiating sand transport. Coastal dune development is a result of the complex interplay of several factors like the type of sediment, the sand influx, and the presence of vegetation. Important aspects for coastal dune formation are the topography of the area, the wave climate and, on a larger scale, sea level \citep{Hesp_2002}. With heights that can vary between a few meters to over 30 m, dunes represent not only tourist attraction in coastal zones, but they also work as barriers against waves, act as sand reserve for beaches and protect the land behind them from erosion and salt intrusion. Understanding the morphology and dynamics of coastal dunes is thus of interest also for federal, state, and local government units responsible for the protection and management of coastal dune areas \citep{IBAMA_2003}. 

For strong unidirectional winds blowing from the sea onto the continents, transverse dunes appear along the coasts, moving landward. Barchanoids and transverse dunes extending over several kilometers form the characteristic pattern of the biggest coastal dune field in Brazil, known as ``Len\c{c}\'ois Maranhenses'' \citep{Goncalves_et_al_2003}. Such scenario is typical of {\em{transgressive}} dune fields \citep{Hesp_et_al_1989}, which appear on high energy beaches experiencing high littoral drift. In this work, we present field measurements performed on a sequence of transverse dunes situated at the beginning of the Len\c{c}\'ois Maranhenses. The height profile of seven transverse dunes was measured and compared with numerical simulations using a dune model in two dimensions \citep{Schwaemmle_and_Herrmann_2004}, where in the present work a phenomenological relation for the separation bubble has been introduced to the model, based on the field data for interdune spacing as a function of the dune height and crest shape. 

% ******************** SECTION 2: AREA OF INVESTIGATION *********************

\section{Area of investigation}

\begin{figure} 
\begin{center} 
\includegraphics*[width=0.95\columnwidth]{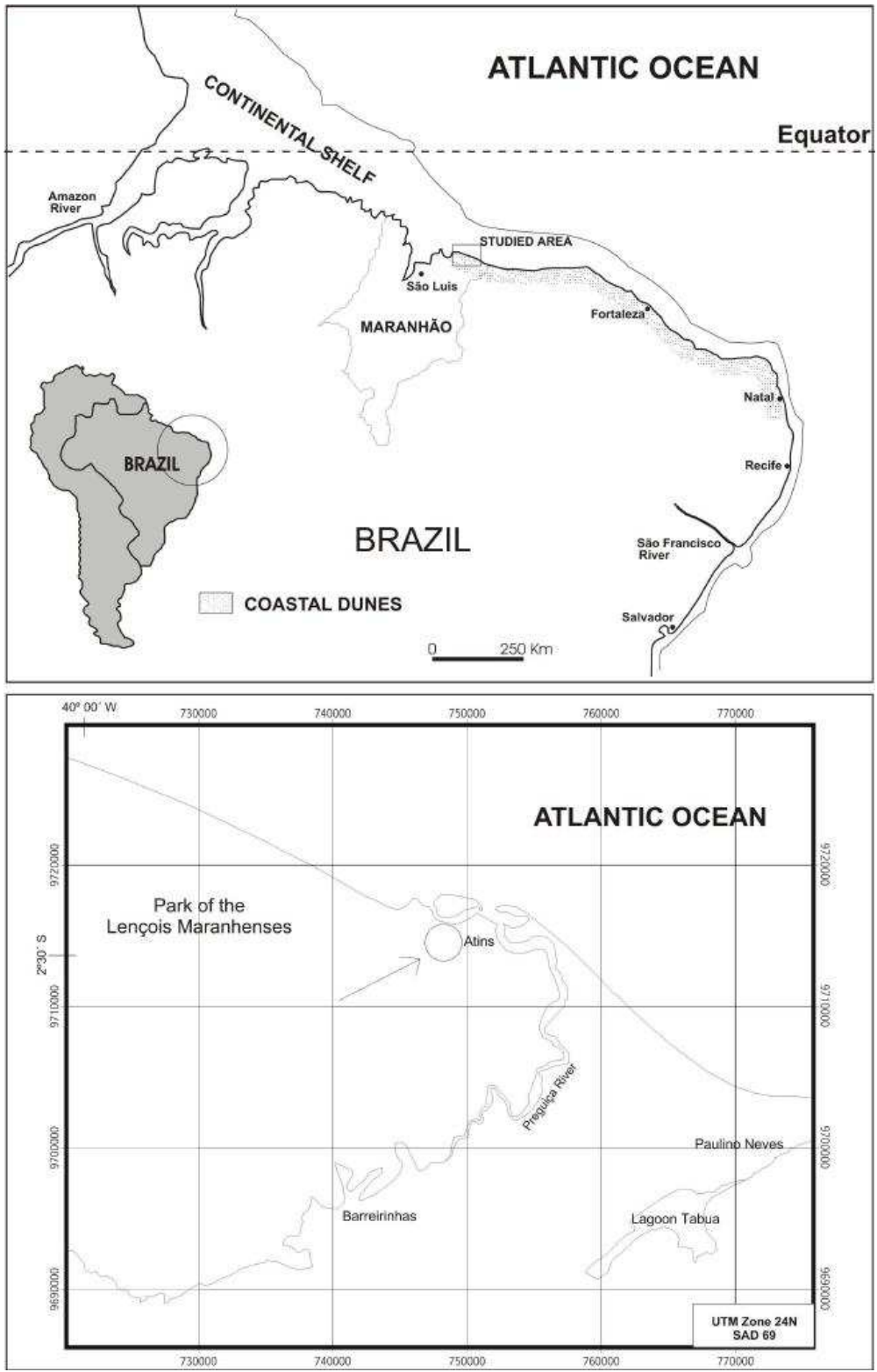} 
\caption{Map of the Len\c{c}\'ois Maranhenses region. The studied transverse dune field is located near the village of Atins, MA.} 
\label{fig:Mapa} 
\end{center} 
\end{figure}

\begin{figure}
\begin{center}
\includegraphics*[width=0.7\columnwidth]{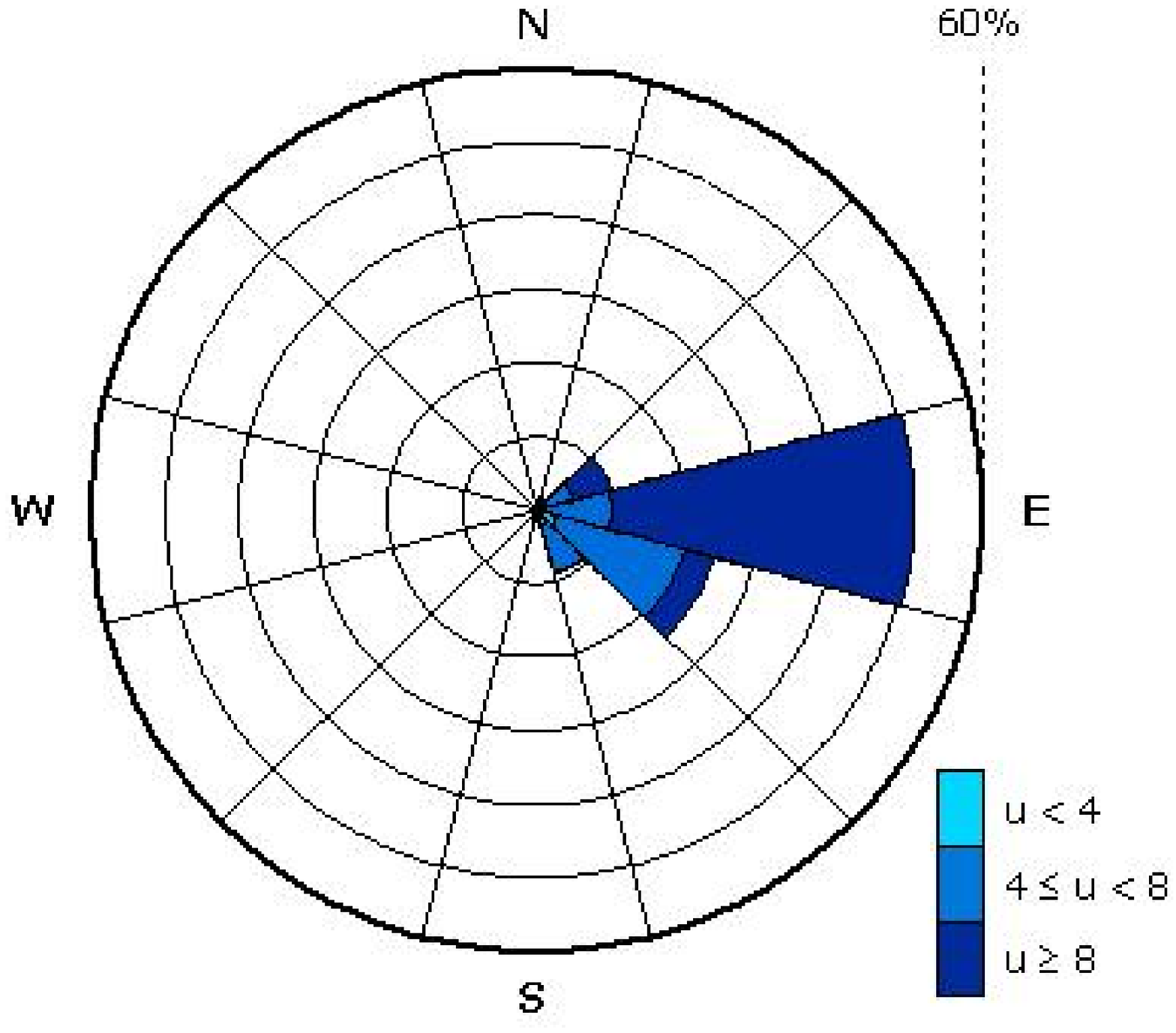}
\caption{Wind rose for the region of the Len\c{c}\'ois Maranhenses corresponding to the period from January to December of 2003. We see that the wind blows mainly from the East. The velocity of the wind (``u'') is shown in units of m$/$s.}
\label{fig:windrose}
\end{center}
\end{figure}

The National Park of the Len\c{c}\'ois Maranhenses, also known as the ``Brazilian Sahara'', is located on the coastal area of the Maranh\~ao State, in northeastern Brazil (Fig. \ref{fig:Mapa}), delimited by the coordinates S $02^{\circ}19'$ and $02^{\circ}45'$, and W $42^{\circ}44'$ and $43^{\circ}29'$. 
It has a total area of 155 thousand hectares, and a coast of a length of 50 km, along which strong winds, reaching velocities of 70 km/h, transport sand from the Atlantic Ocean onto the continent. The wind rose in Fig. \ref{fig:windrose} shows that winds blow mainly from the East.

\begin{figure}
\begin{center}
\includegraphics*[width=0.75\columnwidth]{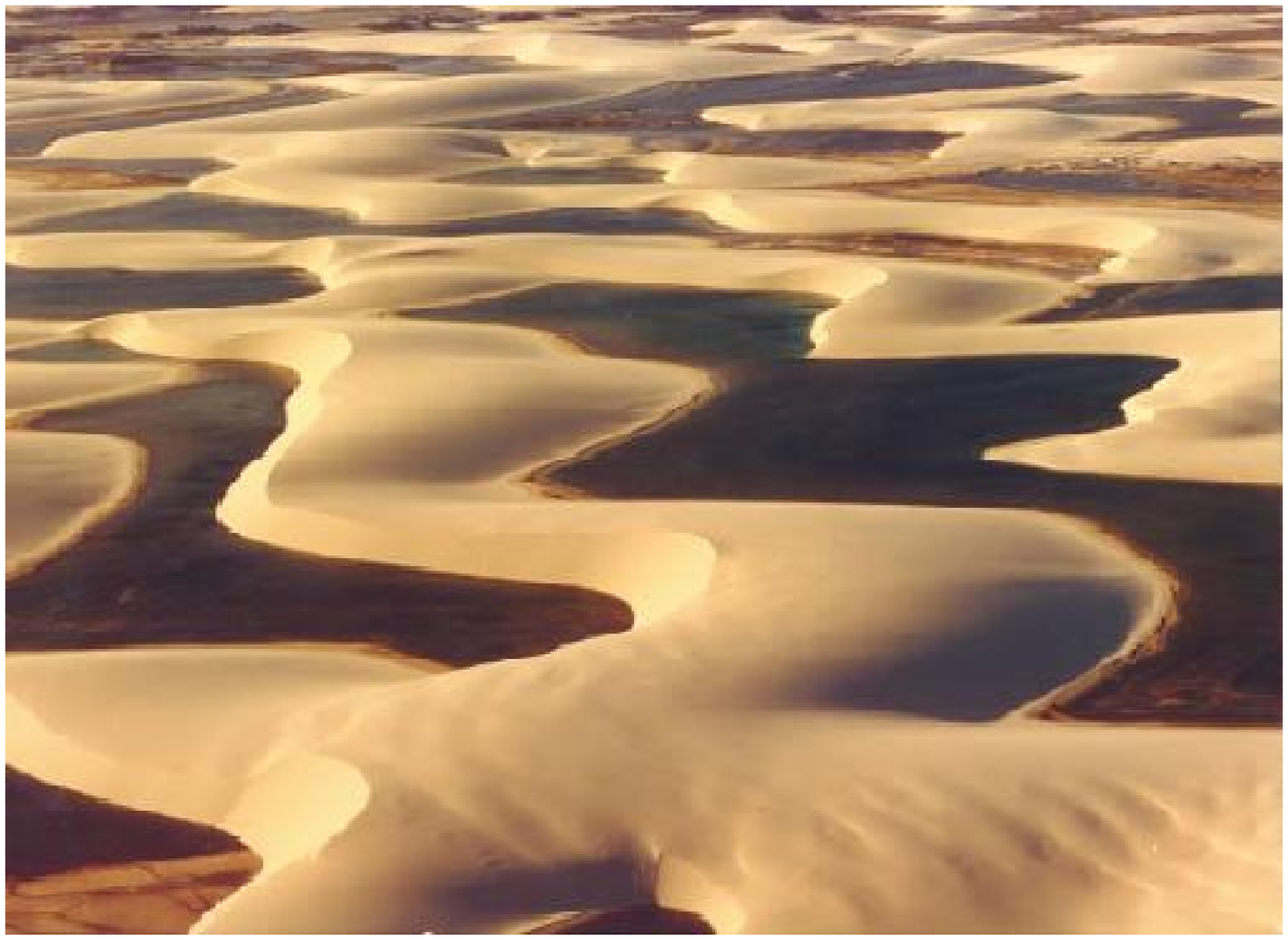}
\caption{Typical landscape of the Len\c{c}\'ois Maranhenses. Chains of barchanoids are separated by interdune lakes that emerge in the wet seasons. Photo: Morais Brito.}
\label{fig:barchanoide}
\end{center}
\end{figure}

\begin{figure}
\begin{center} 
\includegraphics*[width=1.00\columnwidth]{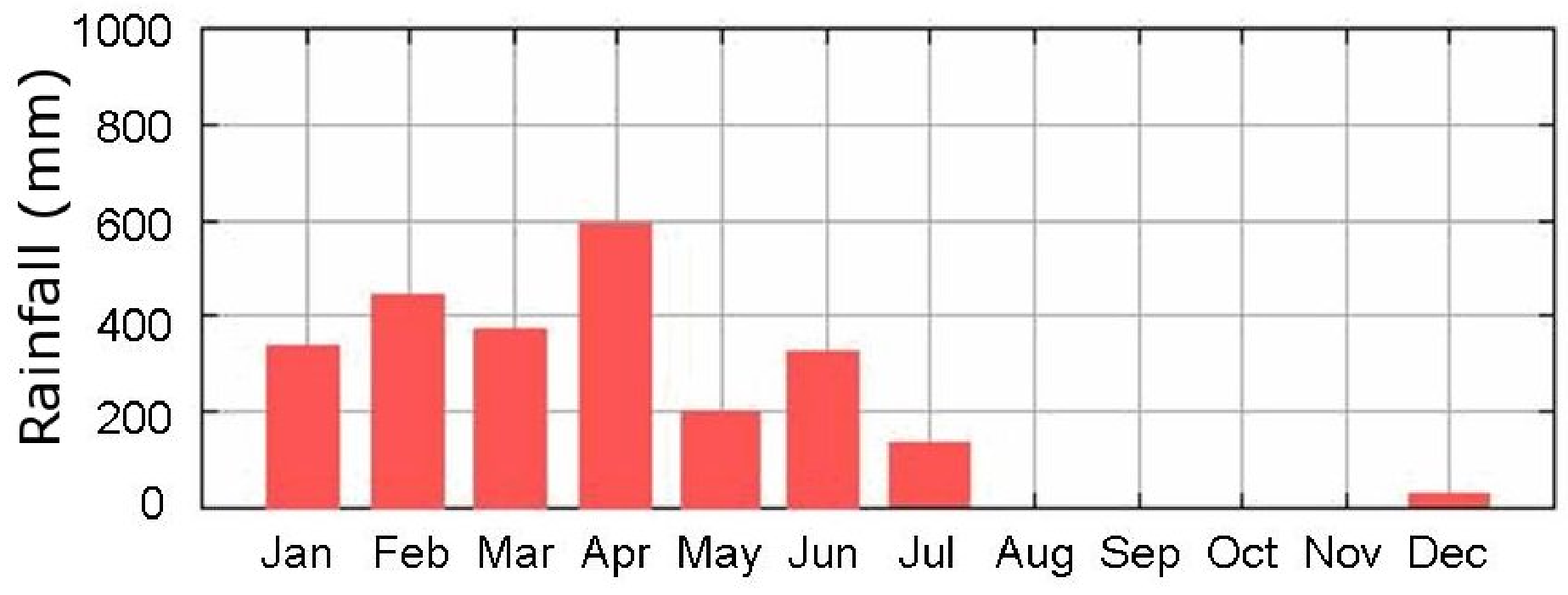}
\caption{Rainfall throughout the year. Vertical axis is in mm month$^{-1}$.}
\label{fig:rain}
\end{center}
\end{figure}

The area of the ``Len\c{c}\'ois'' (or `sheets' in Portuguese) is mainly characterized by the presence of barchanoids and transverse sand dunes separated in the rainy season by lakes and lagoons (Fig. \ref{fig:barchanoide}). 
The area has a semi-humid tropical climate, with little vegetation. The local temperatures are typically 30 to $38^{\circ}$C, with annual rain between 1500 mm and 2000 mm, which is an amount quite different from the mean rainfall in desert areas (in general less than 250 mm). The rainfall distribution, however, is very concentrated, as we can see in Fig. \ref{fig:rain}. In the first months of the year, high rainfall makes sand transport difficult, thus implying low dune mobility. A large fraction of the interdune lakes formed in this period disappear in the dry season. The landscape of the dune field appears to change continuously, indeed dune mobility implies that the lakes often reappear in different places with different contours. The lowest rainfall indices in the studied area refer to the period between August and November. The field work has been carried out from 23 to 29 of September of 2003, in the middle of the dry season. 

\begin{figure}
\begin{center}
\includegraphics*[width=0.8\columnwidth]{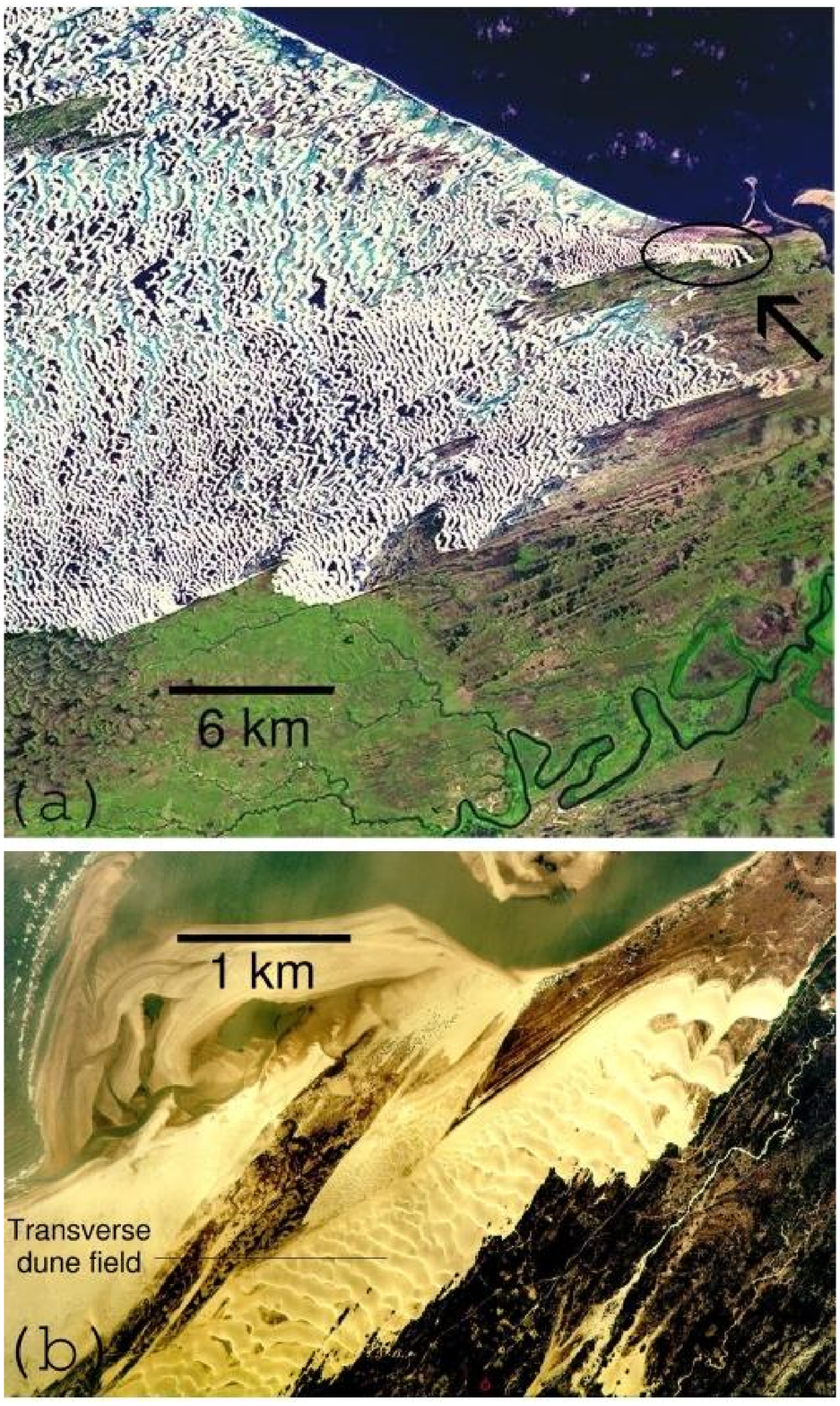}
\caption{In (a) we show a Landsat image (1984) of the Len\c{c}\'ois Maranhenses, where the studied transverse dune field is indicated by the arrow. In (b) we show an image of the studied field, from Embrapa Monitoramento por Sat\'elite, 1999.}
\label{fig:field}
\end{center}
\end{figure}

The transverse dune field studied is shown in Fig. \ref{fig:field}. It is situated in the vicinity of Atins, a small village in the Maranh\~ao State (MA), just 1 km from the sea. These dunes have a height of typically 10 m and are closely spaced, separated by a few meters only, on a dense sand sheet in the beginning of the Len\c{c}\'ois Maranhenses. They may be classified as ``transgressive dunes'', which are defined in most of the cases as dunes that {\em{develop}} on sand sheets \citep{Hesp_et_al_1989}. 

Near to the transverse field, several small barchans and barchanoids are found, with heights $\sim 50$ cm $-$ 1 m. These smaller dunes emerge on a flat surface covered with less sand, extending from the beach up to the transverse dune field, as shown in Fig. \ref{fig:field}(b). The size difference between the two neighbouring dune groups is clearly visible. The dune field studied here represents the genesis of the Len\c{c}\'ois, where the characteristic transverse profile originates. The dark areas in Fig. \ref{fig:field}(b) correspond to low vegetation, called ``mangues''. The coastal zone near Atins is also characterized by a high concentration of plants developed on parabolic dunes of different sizes. Vegetation appears due to humidity from the ocean and interdune lakes, and grows during the wet seasons, appearing also, although more rarely, on some parts of transverse dunes. 

In the next section we present the results of our measurements. In Section 4, we describe the dune model originally introduced in \cite{Schwaemmle_and_Herrmann_2004} and propose a modification of the separation bubble to simulate closely spaced transverse dunes, based on the data of our measurements of dune spacing and crest-brink distances of the dunes. Comparison of the results of our simulations with the measured profile and discussions are presented in Section 5. Finally, in Section 6 we present our conclusions.

% *********************** SECTION 3: MEASUREMENTS  ***************************

\section{Measurements}

\begin{figure}
\begin{center}
\includegraphics*[width=1.0\columnwidth]{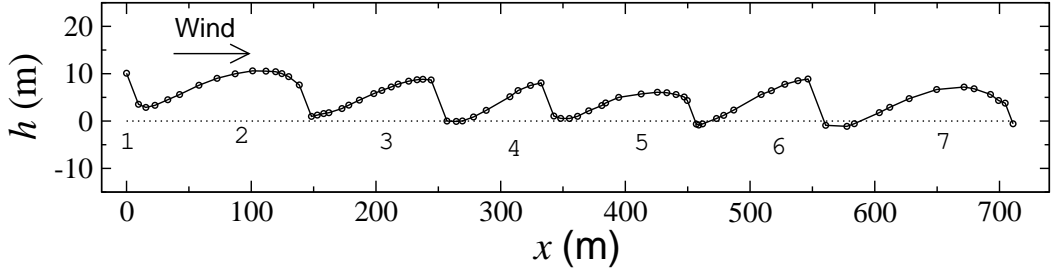}
\caption{Height profile of the transverse dune field studied.}
\label{fig:profile}
\end{center}
\end{figure}

\begin{figure}
\begin{center}
\includegraphics*[width=0.6\columnwidth]{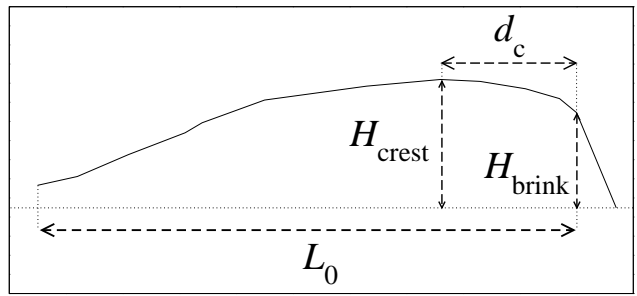}
\caption{Profile of dune 5 in Fig. \ref{fig:profile} with definition of variables $H_{\mathrm{crest}}$, $H_{\mathrm{brink}}$, $d_{\mathrm{c}}$ and $L_0$ in Table \ref{tab:measurements}. Heights ($H$) are measured relative to the foot ot the slip face.}
\label{fig:sketch}
\end{center}
\end{figure}

A tachimeter was used to measure the height variations at 77 different points along the surface of the dunes, which define a profile with a length of almost 720 m, as shown in Fig. \ref{fig:profile}. To interpolate the entire profile of the transverse dune field, we used the relative distances and height differences between the measured points and a fixed angle ${\theta}_{\mathrm{c}} = 34^{\circ}$ for the slip face for each dune. The reference height $h=0$ is set at the foot of the slip face of dune 3, whereas $x=0$ corresponds to the brink of dune 1, for which we did not measure the whole profile. Our measurements have error estimates of $0.1$ m for the height coordinate and $1$ m for the horizontal axis. 
The measured transverse dunes have heights of typically $7-10$m and a crest-to-crest distance, $D_{\mathrm{cc}}$, of $90-130$ m, as shown in Table 1. The sixth column of this table shows the aspect ratio $r$ of these dunes, which is obtained by dividing their heights at the crest, $H_{\mathrm{crest}}$, by the corresponding values of the length of the windward side, $L_0$, and we find that $r = 0.08 \pm 0.02$ is consistent for all dunes. The quantities listed in Table 1 are defined in Fig. \ref{fig:sketch}.
\begin{table}
\begin{center} 
\begin{tabular}{|c|c|c|c|c|c|c|}
\hline
\hline
Dune & $H_{\mathrm{crest}}$(m) & $H_{\mathrm{brink}}$(m) & $d_{\mathrm{c}}$(m) & $L_0$(m) & $r = H_{\mathrm{crest}}/L_0$ & $D_{\mathrm{cc}}(i,i+1)$(m) \\ \hline \hline
1 & $-$ 	   & $6.6 \pm 0.1$ & $-$ 	  & $-$		 & $-$ 		 & $-$ \\ \hline
2 & $9.6 \pm 0.1$ & $6.6 \pm 0.1$ & $32 \pm 1$ & $123 \pm 1$ & $0.078 \pm 0.005$ & $132 \pm 1$  \\ \hline
3 & $8.8 \pm 0.1$ & $8.6 \pm 0.1$ & $7  \pm 1$ & $96  \pm 1$ & $0.092 \pm 0.005$ & $94  \pm 1$ \\ \hline
4 & $7.0 \pm 0.1$ & $7.0 \pm 0.1$ & $0  \pm 1$ & $65  \pm 1$ & $0.108 \pm 0.005$ & $95  \pm 1$ \\ \hline
5 & $6.9 \pm 0.1$ & $4.9 \pm 0.1$ & $24 \pm 1$ & $95  \pm 1$ & $0.073 \pm 0.005$ & $119 \pm 1$ \\ \hline
6 & $9.7 \pm 0.1$ & $9.7 \pm 0.1$ & $0  \pm 1$ & $88  \pm 1$ & $0.110 \pm 0.005$ & $124 \pm 1$ \\ \hline
7 & $7.9 \pm 0.1$ & $4.3 \pm 0.1$ & $35 \pm 1$ & $127 \pm 1$ & $0.062 \pm 0.005$ & $-$ \\ \hline
\end{tabular}
\end{center}
\vspace{0.5cm}
\caption{For each dune in the profile of Fig. \ref{fig:profile} we identify in columns 2, 3, 4, 5 and 6, respectively: The dune height at the crest, $H_{\mathrm{crest}}$, the height at the brink, $H_{\mathrm{brink}}$, the crest-brink distance $d_{\mathrm{c}}$, the windward side length $L_0$ and the aspect ratio $r$. The seventh column shows the crest-to-crest distance $D_{\mathrm{cc}}(i,i+1)$ between the $i-$th and $(i+1)-$th dunes. $H_{\mathrm{crest}}$, $H_{\mathrm{brink}}$, $d_{\mathrm{c}}$ and $L_0$ are shown in Fig. \ref{fig:sketch} for dune 5.}\label{tab:measurements}
\end{table}

It is surprising that although the dunes have more or less the same height, the position of the brink with respect to the crest, $d_{\mathrm{c}}$, varies strongly. This is in strong contrast with the situation of single barchans, for which it has been observed from field measurements in southern Morocco that the brink gets closer to the crest for higher barchan dunes \citep{Sauermann_et_al_2000}. In addition, differences between small and large barchans have been recently reported by \cite{Hesp_and_Hastings_1998}. On the other hand, \cite{Long_and_Sharp_1964} have more generally identified two kinds of barchan dunes: the ``slim'' and ``fat'' barchans, where the latter dunes may have been the more ``morphologically complex and areally larger dune masses'' formed by coalescence and fusion, and behave in a different way as the slim ones. \cite{Bagnold_1941} regarded this phenomenon as the product of the lag in-between changes of wind conditions and the resulting changes in sand movement.

\begin{figure}
\begin{center}
\includegraphics*[width=0.95\columnwidth]{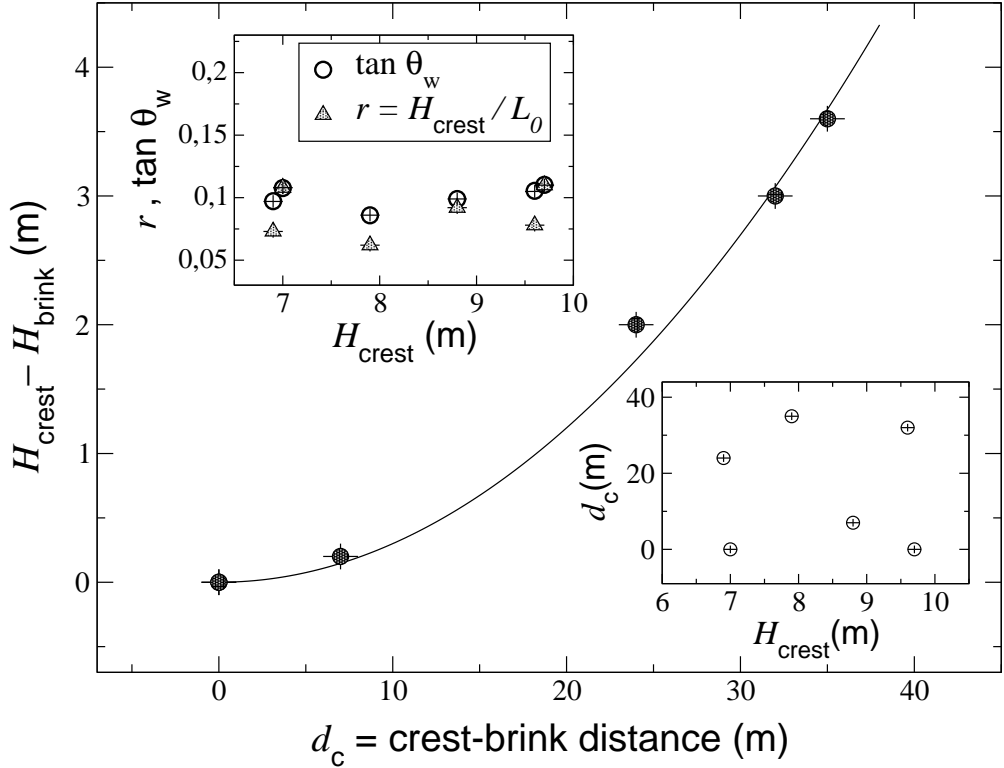}
\caption{Difference between dune heights at the crest and at the brink, as a function of the distance $d_{\mathrm{c}}$ between the referred two points for the measured dunes. For dunes 4 and 6 the crest coincides with the brink, while the corresponding data for dune 1 were not recorded. The observed quadratic relation is a consequence of the parabolic shape of the dune at the crest. The upper inset shows the aspect ratio $r = H_{\mathrm{crest}}/{L_0} = 0.08 \pm 0.02$ (triangles) and the average windward side inclination $\tan{{\theta}_{\mathrm{w}}} = H_{\mathrm{crest}}/{(L_0-d_{\mathrm{c}})} = 0.10 \pm 0.02$ (circles), ${\theta}_{\mathrm{w}} \approx 6^{\circ}$. The lower inset shows the crest-brink distance versus the height of the crest.} 
\label{fig:Hbrink-Hcrest}
\end{center}
\end{figure}

\begin{figure}
\begin{center}
\includegraphics*[width=0.95\columnwidth]{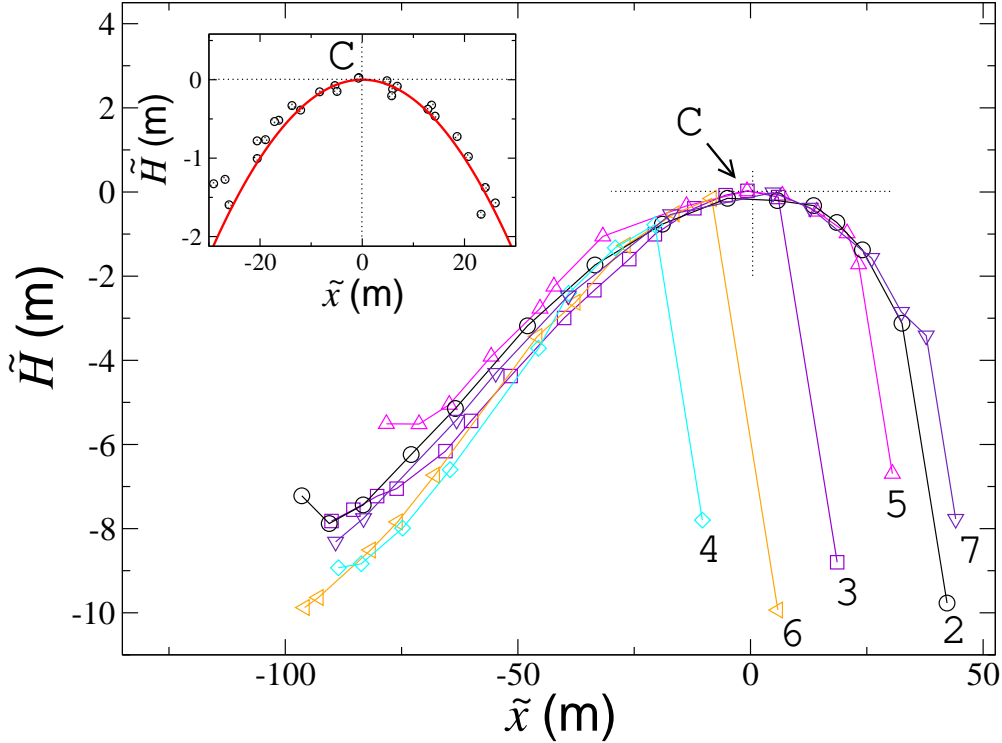}
\caption{The main plot shows the height profile of dunes $2-6$, where the dunes are identified by numbers at their slip faces, and are represented by different symbols. The windward sides of these dunes are shown superposed to each other in such a way to form a parabola with maximum at $C(0,0)$. This behaviour is more clearly shown in the inset, where the circles represent the height profile of all dunes of the main plot, and are fitted by a parabola (continuous line). The symbol ``$\sim$'' over $H$ and $x$ indicates that these variables are measured, both in the inset and main plot, relative to point $C$ for each dune. The dotted lines in the inset and main plot are guides to the eye meant to localize the maximum $C$ of the parabola.} 
\label{fig:parabola}
\end{center}
\end{figure}

The upper inset of Fig. \ref{fig:Hbrink-Hcrest} shows the inclination ${\theta}_{\mathrm{w}}$ of the windward side (circles) together with the aspect ratio $r$ (triangles) of the measured transverse dunes. Here ${\theta}_{\mathrm{w}}$ is defined through the relation $\tan{{\theta}_{\mathrm{w}}} = {H_{\mathrm{crest}}}/{(L_0-{d_{\mathrm{c}}})}$, where $d_{\mathrm{c}}$ is the horizontal distance measured from the crest to the brink of the dune. 
From the value $\tan{{\theta}_{\mathrm{w}}}=0.10 \pm 0.02$, we get ${\theta}_{\mathrm{w}} \approx 6^{\circ}$. The main plot of Fig. \ref{fig:Hbrink-Hcrest} shows that the difference between the heights at the crest and at the brink is essentially quadratic in $d_{\mathrm{c}}$. Moreover, Fig. \ref{fig:parabola} shows that the transverse dunes present a nearly parabolic shape at their crest. In this figure, we lump the profiles of dunes $2-6$ together, and we can see that the windward sides of all dunes at their crests superimpose in a single parabolic curve with maximum at point $C$. The vertical (horizontal) axis of the main plot gives the profile $\tilde{H}$ (horizontal distance $\tilde{x}$) measured relative to point $C$ for all dunes. The empty circles in the inset of Fig. \ref{fig:parabola} represent the sampled profiles for all dunes close to point $C$, as indicated by the dotted lines, while the continuous line corresponds to the curve $y=-k_1x^2$, with $k_1= 0.0025$.

For dune 3, we performed correlated measurements of the wind velocity $v$ and the sand flux $q$ on two different points along its windward side. This dune has a height $H_{\mathrm{crest}}=8.8$ m measured from the foot of its slip face up to the crest, and the length of its windward side is approximately 100 m. We placed an anemometer at a height of 1m over the ground, and the registered wind velocity was averaged over time intervals of 10 min. An average wind velocity of $\left<v_{\mathrm{A}}\right>=7.5 \pm 1.2$m/s was registered at the brink of the dune (point ``A''). Then, we moved the anemometer to a point at about the middle of the windward side (point ``B''), and the average wind velocity registered there was $\left<v_{\mathrm{B}}\right>=5.4 \pm 0.5$m/s. A second anemometer, placed at one point outside of the dune field, at about 10 m from the transverse dunes on a sand sheet where a field of small barchans developed close to the ocean, registered an average wind velocity of ${\left<v_{\mathrm{r}}\right>} = 5.3 \pm 0.7$ m/s, also at a height of 1 m. We did not measure the wind velocity at different heights over the ground, and thus we could not estimate the shear velocity $u_{{\ast}0}$ and the aerodynamic roughness $z_0$ of the surface. This quantities can be obtained from the logarithmic profile of the wind velocity $v(z)$ \citep{Pye_and_Tsoar_1990}:
\begin{equation}
v(z)={\frac{u_{\ast}}{\kappa}}{\ln{\frac{z}{z_0}}}, \label{eq:log_profile}
\end{equation}
where $z$ is the height over the ground and $\kappa = 0.4$ is the von K\'arm\'an constant, and $u_{\ast}$ is used to define the shear stress of the wind:
\begin{equation}
\tau = {\rho}_{\mathrm{air}}{u_{\ast}^2}, \label{eq:shear_stress}
\end{equation}
where ${\rho}_{\mathrm{air}}$ is the density of the air. If we take, for example, a roughness length $z_0 = 2.5 \times 10^{-3}$ m, we then obtain the undisturbed shear velocity over the plane $u_{{\ast}0} \approx 0.36$ m/s, using $v(z = 1{\mbox{m}}) = 5.35$ m$/$s. It is interesting to note that these were essentially the same values of $u_{{\ast}0}$ and $z_0$ found in a recent work on the dune field of Jericoacoara, also in northeastern Brazil \citep{Sauermann_et_al_2003}. Since we could not measure accurately the values of these quantities, we use in our simulations (see Section 5) estimated values of $u_{{\ast}0}$ close to $0.4$ m/s and $z_0$ between $0.8-3$ mm.

\begin{figure}
\begin{center}
\includegraphics*[width=0.7\columnwidth]{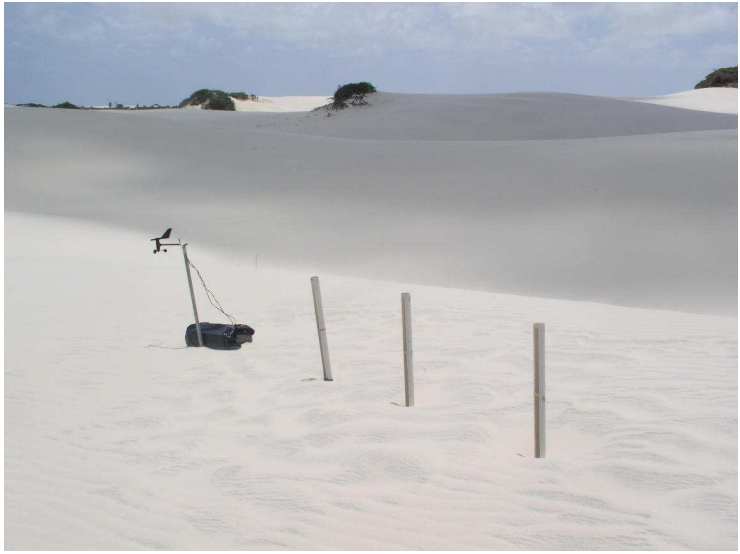}
\caption{Measuring wind velocity and sand flux using an anemometer and the three traps at the different points of dune 3 in Fig. \ref{fig:profile}.}
\label{fig:traps}
\end{center}
\end{figure}

At the same positions of the dune where the anemometer was placed, as shown in Fig. \ref{fig:traps}, we fixed 3 cylindrical traps to measure the corresponding sand flux. Each trap consists of a cylinder of 1 m height and 4.8 cm diameter, and has an open vertical slot of 1 cm width. 
The traps were turned with their slit facing the wind and their bottom buried in the sand (see Fig. \ref{fig:traps}). The back of the traps has an opening of 2 cm width covered by a permeable fabric with pores smaller than the grain diameter. Initially, the traps were placed with their back showing in the wind direction. Then, they were simultaneously turned to align their openings with the wind direction and sand was collected, in a similar way as in \cite{Sauermann_et_al_2003}. From the mass $m$ of the collected sand, the width $w$ of the opening and the time $T$ necessary to fill the traps, we calculated the sand flux $q=m/(Tw)$, for which we use $m={\pi}r^2h_c \times {\rho}_{\mathrm{quartz}}$, with $r=2.4$ cm and ${\rho}_{\mathrm{quartz}} = 2650$ kg/m$^3$. For each point two traps were used which captured grains travelling up to a height $h_c \approx 17$ cm and another one at a height of $1.5$ cm above the ground. The mean values $\left<{q}\right>$ obtained from two runs using these traps are, for points ``A'' (brink) and ``B'' (middle of windward side) respectively, $\left<{q_{{\mathrm{A}}}}\right> = 0.023 \pm 0.005$ kg/m$\cdot$s and $\left<{q_{{\mathrm{B}}}}\right> = 0.009 \pm 0.001$ kg/m$\cdot$s, where the error estimates are defined by the maximum and mininum values registered for $q$ at each point, as in the case of the measurements of the wind velocity. A rigorous treatment of our measurements of sand flux would require reference to a recent work on sampling efficiency of a vertical array of aeolian sand traps \citep{Li_and_Ni_2003}. Since we have only two measurement points for wind and flux, we report here the measured values but in this work we do not show comparison with model predictions for these data. We present typical simulations of the transverse profile and compare our results with the observed dune shape.

% ****** SECTION 4: A TWO-DIMENSIONAL MODEL FOR TRANSVERSE DUNES *************

\section{A two-dimensional model for transverse dunes}

A two-dimensional model including the main processes of dune morphology was recently introduced \citep{Kroy_et_al_2002} and extended to study transverse dunes in a field \citep{Schwaemmle_and_Herrmann_2004}. From this model, the air shear stress $\tau$ and the saltation flux $q$ at the ground are numerically calculated in each iteration, as well as the flux due to avalanches at the slip face of the dunes. The shear stress perturbation is calculated in the two-dimensional Fourier space using the algorithm of \cite{Weng_et_al_1991} for the components ${\tau}_x$ and ${\tau}_y$:
\begin{equation}
{{\hat{\tau}}}_x(k_x,k_y)={\frac{2\,h(k_x,k_y){k}_{x}^2}{|k|\,U^2(l)}}\left({1+\frac{2\,{\ln(L|k_x|) + 4{\gamma} + 1 + {\mbox{i}}\,{\mbox{sign}}(k_x){\pi}}}{\ln{\left({l/z_0}\right)}}}\right), \label{eq:tau_x}
\end{equation}
and
\begin{equation}
{{\hat{\tau}}}_y(k_x,k_y)={\frac{2\,h(k_x,k_y)k_xk_y}{|k|\,U^2(l)}}, \label{eq:tau_y}
\end{equation}
where the axis $x$ ($y$) points parallel (perpendicular) to the wind direction, $k_x$ and $k_y$ are wave numbers, $|k|=\sqrt{k_{x}^2+k_{y}^2}$, $\gamma = 0.577216$ (Euler's constant), $z_0$ is the roughness length and $L$ is the characteristic length for a hill \citep{Hunt_et_al_1988}, and is defined as the length measured from the half height of the windward side to the crest. $U(l)=v(l)/v(h_{\mathrm{m}})$ is the undisturbed logarithmic profile (\ref{eq:log_profile}) calculated at height $l$, which is given by 
\begin{equation}
l={\frac{2{\kappa}^2L}{\ln{{l}/z_0}}}, \label{eq:l}
\end{equation}
and normalized by the velocity at a reference height $h_{\mathrm{m}} = L/{{\sqrt{{\log{L/z_0}}}}}$, which separates the middle and upper flow layers defined in the model by \cite{Hunt_et_al_1988}. It has been verified from simulations that an open system of transverse dunes seems to reach lateral invariance under unidirectional wind \citep{Schwaemmle_and_Herrmann_2004}. Although this observation is not strictly valid in the field due to the multi-directional nature of secondary airflow patterns \citep{Walker_and_Nickling_2002}, the assumption yields to important advantages in transverse dune field calculations. Thus, for modelling transverse dune fields, as opposed to the situation of barchan dune modelling \citep{Schwaemmle_and_Herrmann_2005}, we neglect the $y$-coordinate and consider height profile variations in the $x$ direction only, since much less computational time is consumed in this case, and larger dune fields can be simulated. In this way, we use the shear stress perturbation evaluated for a slice of a transverse dune translationally invariant in the $y$ direction \citep{Kroy_et_al_2002}:
\begin{equation}
{{\hat{\tau}}}(x) = f_1(z_0/L)\left({{\frac{1}{\pi}}\,{{\int_{-\infty}^{\infty}}}{\frac{h^{\prime}}{x-\xi}}d{\xi}+f_2(z_0/L)h^{\prime}}\right), \label{eq:shear_stress_perturbation}
\end{equation}
where $h^{\prime}=dh/dx$ is the height derivative in $x$ direction, and the constants $f_1(z_0/L)$ and $f_2(z_0/L)$ depend logarithmically on the ratio $z_0/L$. The shear stress is then given by:
\begin{equation}
\tau = \tau(x)={\tau}_0{[1+{{\hat{\tau}}}(x)]}, \label{eq:shear_stress_unidimensional}
\end{equation}
where ${\tau}_0$ is the undisturbed air shear stress over a flat ground. Of course, the fluctuations in the $y-$direction that can be observed in Fig. \ref{fig:field}(b) may be of further importance for the dune dynamics, but are not considered in this approach. The sand flux $q$ is calculated as: 
\begin{equation}
q = {\rho}u, \ \  \ \ {\frac{{\partial}{\rho}}{{\partial}{t}}} + {\frac{{\partial}}{{\partial}{x}}}{({\rho}{u})} = \frac{1}{T_s}{\rho}\left({1-\frac{\rho}{{\rho}_s}}\right) 
\label{eq:sand_flux},
\end{equation}
where ${\rho}$ and $u$ are, respectively, the mean density and the mean velocity of the saltating grains, and are both integrated in vertical direction. The saturated density at the steady state, ${\rho}_s$, and the characteristic time $T_s$ that defines the transients of the flux are given by:
\begin{equation}
{\rho}_s=\frac{2{\alpha}}{g}({\tau}-{\tau}_t) \ \ \mbox{and} \ \ T_s = \frac{2{\alpha}{u}}{g}{\frac{{\tau}_t}{{\gamma}({\tau}-{\tau}_t)}} \ ,
\end{equation}
where ${\tau}_{\mathrm{t}}$ is the threshold shear stress for saltation transport \citep{Bagnold_1941}, and $\alpha$ is a model parameter that represents an effective restitution coefficient for the grain-bed interaction. The vertical component of the initial velocity of the ejected grains after the impact of saltating grains onto the bed (``splash'') is equal to the average gain in horizontal velocity of the impacting grain, after its acceleration by the wind, multiplied by $\alpha$ \citep{Sauermann_et_al_2001}. On the other hand, $\gamma$ is the second model parameter, to which the erosion rate is proportional \citep{Sauermann_et_al_2001}.
 
The steady state is assumed to be reached instantaneously, since it corresponds to a time scale several orders of magnitude smaller than the time scale of the surface evolution. Thus, time-dependent terms are neglected. Furthermore, the velocity $u$ of the grains in the saltation layer is modelled through the following equation:
\begin{equation}
\frac{3}{4}C_d{\frac{{\rho}_{\mathrm{air}}}{{\rho}_{\mathrm{quartz}}}}{\frac{1}{d}}({v}_{\mathrm{eff}} - u)|{v}_{\mathrm{eff}}-u| - {\frac{g}{2{\alpha}}}  = 0, \label{eq:u}
\end{equation}
where:
\begin{equation}
{v}_{\mathrm{eff}} = {\left({{\frac{2}{\kappa}}{\sqrt{{\frac{z_1}{z_m}}{u_{\ast}^2} + \left({1-{\frac{z_1}{z_m}}}\right)u_{{\ast}{\mathrm{t}}}^2}} + {\left({\ln{\frac{z_1}{z_0^{\mathrm{sand}}}} - 2}\right)}\frac{u_{{\ast}{\mathrm{t}}}}{\kappa}}\right)}, \label{eq:v_eff}
\end{equation}
and $u_{{\ast}{\mathrm{t}}} = \sqrt{{{\tau}_{\mathrm{t}}}/{\rho}_{\mathrm{air}}}$. In Eq. (\ref{eq:v_eff}) $v_{\mathrm{eff}}$ is an effective velocity calculated at a reference height $z_1$ within the saltation layer between the roughness length $z_0^{\mathrm{sand}}$ associated with the grain diameter and the mean saltation height $z_{\mathrm{m}}$. The value of $z_0^{\mathrm{sand}}$ is smaller than the aerodynamic roughness \citep{Pye_and_Tsoar_1990}, and is taken as of the order of $d/10$, where $d$ is the grain diameter. The velocity $u$ is calculated from $v_{\mathrm{eff}}$ using eq. (\ref{eq:u}). Typical values of the model parameters $z_1$, $\alpha$ and $\gamma$ defined above have been determined from comparisons of the model with microscopic simulations and wind tunnel measurements of sand flux and saturation time transients \citep{Sauermann_et_al_2001}. We refer to these results and use the same values for the constants and microscopic model parameters as in previous work \citep{Sauermann_et_al_2001,Schwaemmle_and_Herrmann_2004}: $g=9.81$ m/s$^2$, $\kappa =0.4$, ${\rho}_{\mathrm{air}}=1.225$ kg/m$^{3}$, ${\rho}_{\mathrm{quartz}}=2650$ kg/m$^{3}$, $z_1 = 0.005$ m, $z_{\mathrm{m}} = 0.04$ m, $\alpha = 0.35$, ${\gamma}=0.4$, ${u}_{{\ast}{\mathrm{t}}} \simeq 0.3$ m/s, $d=250$ ${\mu}$m and $C_d = 3$. It has to be emphasized that these values have not been obtained from measurements on the field studied in the present work. Of course, a detailed quantitative description of the sand flux and wind velocity over the transverse dunes investigated here would require more field data. Furthermore, the shape and dynamics of dunes may be strongly influenced by the sorting and size distribution of grains, which is in general heterogeneous rather than well defined at one single size, and is also related to the dune's age \citep{Besler_2002}. In our model, such a feature is not taken into account. A recent study has shown that the mean grain diameter of the sand in the area investigated in the present work lies between $177$ and $354$ $\mu$m \citep{IBAMA_2003}. In simulations, a grain diameter of $d=250$ ${\mu}$m is typically used, around which the grain size distribution of sand dunes appears sharply peaked as shown by \cite{Pye_and_Tsoar_1990}, thus defining a threshold shear velocity of about $0.3$ m/s \citep{Sauermann_et_al_2001,Sauermann_et_al_2003}. 

For the avalanches at the slip face, the BCRE model \citep{Bouchaud_et_al_1994} is used to calculate the new surface after the relaxation by sand transport downwind: 
\begin{equation}
\frac{{\partial}h}{{\partial}t} = -C_aR{\left({\frac{{\partial}{h}}{{\partial}{x}} - \tan {\Theta}}\right)}
\end{equation}
\begin{equation}
\frac{{\partial}R}{{\partial}t} + {\frac{{\partial}}{{\partial}x}}{Ru_a} = C_aR{\left({\frac{{\partial}{h}}{{\partial}{x}} - \tan {\Theta}}\right)},
\end{equation}
where $h$ is the height of the non-moving sand bed, $R$ is the height of the moving layer, $C_a$ is a model parameter and ${u_a}$ is the velocity of the grains in the moving layer. Avalanches are considered to occur instantaneously, since the time scale associated with them is negligible if compared with the time scale of the surface evolution \citep{Schwaemmle_and_Herrmann_2004}. 

The evolution of the dune surface is determined by erosion and deposition of sand, and its profile can be obtained from the mass conservation equation:
\begin{equation}
{\partial}_t{\rho} + \frac{{\partial}{\Phi}}{{\partial}x} = 0, 
\end{equation}
where the sand density, $\rho$, and the sand flux per time unit and area, ${\Phi}$, are integrated over the vertical coordinate assuming that the dune has a constant density of ${\rho}_{\mathrm{sand}}$:
\begin{equation}
h = \frac{1}{{\rho}_{\mathrm{sand}}}\,\int {\rho}dz, \ \ \ \ q = \int {\Phi}dz,
\end{equation}
where ${\rho}_{\mathrm{sand}} = 0.62 {\rho}_{\mathrm{quartz}}$, leading to the following equation for the surface height $h$:
\begin{equation}
\frac{{\partial}h}{{\partial}t} = - {\frac{1}{{\rho}_{\mathrm{sand}}}}{\frac{{\partial}{q}}{{\partial}x}}. \label{eq:time_evolution}
\end{equation}
Equation (\ref{eq:time_evolution}) is used to determine the characteristic time scale of the model, typically $3-5$ hours for every iteration. Simulation steps may be summarized as follows: (i) the shear stress over the initial surface (a sand bed with small fluctuations as described below) is calculated using Eqs. (\ref{eq:shear_stress_perturbation}) and (\ref{eq:shear_stress_unidimensional}); (ii) from the shear stress, the sand flux is calculated with Eq. (\ref{eq:sand_flux}); (iii) the change in the surface height is computed from mass conservation (Eq. (\ref{eq:time_evolution})) using the calculated sand flux; and (iv) if the inclination of the surface gets larger than $34^{\circ}$, avalanches occur and a slip face is developed. Steps (i) $-$ (iv) are iteratively computed, and in case of transverse dunes, the evolution of the dune field depends on the boundary conditions \citep{Schwaemmle_and_Herrmann_2004}. For periodic boundary conditions, it was shown in \cite{Schwaemmle_and_Herrmann_2004} that the dunes increase continuously in height though they keep their average aspect ratio, thus comparison with real profiles are made taking simulation snapshots of similar dune size as observed in the field.

For a dune with slip face, flow separation occurs at the brink, which represents a discontinuity of the surface. The flow is divided into two parts by a streamline connecting the brink with the ground at the reattachment point. These streamlines define the separation bubble, inside which eddies occur and the flow is often re-circulating, and may present a complex multi-directional pattern \citep{Walker_and_Nickling_2002}, whereas the averaged flow outside is laminar. In the area of the separation bubble, we set the shear stress equal to zero, and in our model no sand transport occurs in the bubble. The length of the separation bubble is thus of major importance for interdune spacing. 

\begin{figure}
\begin{center}
\includegraphics*[width=0.9\columnwidth]{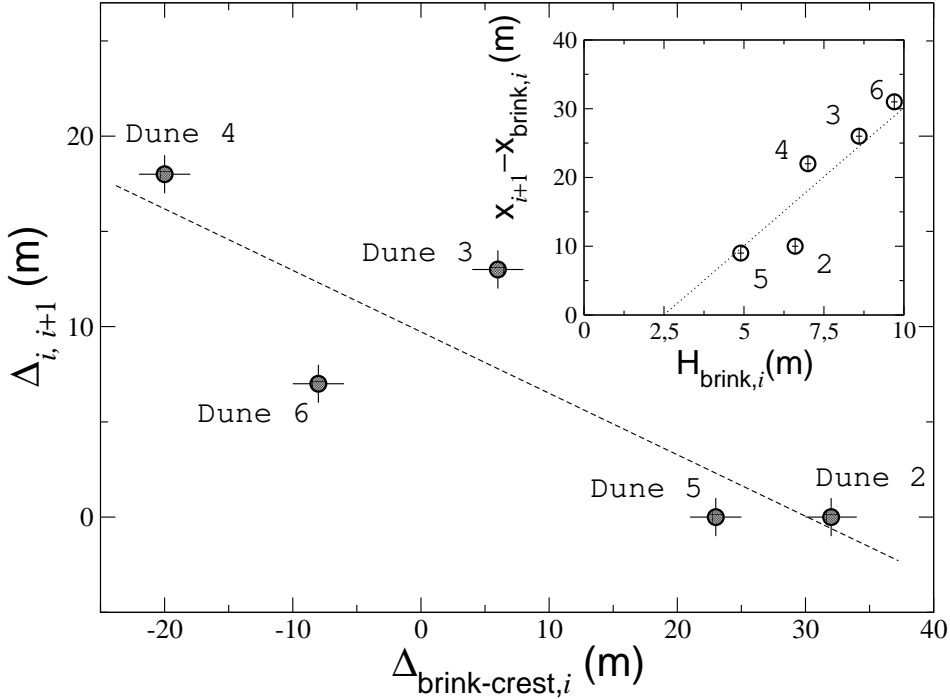}
\caption{Main plot: Distance between the slip face foot of dune $i$ and the windward side foot of dune $i+1$ as a function of the distance measured between the brink and the maximum of the parabola according to Fig. \ref{fig:parabola}. As we can see, the length of the separation bubble appears to decrease linearly with ${\Delta}_{{\mathrm{brink-crest}},i}$. The inset shows the distance $x_{i+1}-x_{{\mathrm{brink}},i}$ between the foot of the windward side of dune $i+1$ and the brink position of dune $i$ as a function of the height of dune $i$ at the brink, $H_{{\mathrm{brink}},i}$. The dotted line in the inset represents the curve $y = 4 \times (H_{{\mathrm{brink}},i} - 2.5)$, which is a guide to the eye meant to show that interdune distances are below four times the dune height at the brink.}
\label{fig:interdune_distances}
\end{center}
\end{figure}

We observed in the field that the net interdune sand flux in the wind direction was zero from the foot of the upwind slip face up to the beginning of the windward side of the downwind dune. If we measure the horizontal distance between the brink of dune $i$ and the point at which the windward side of dune $i+1$ begins, i.e. $x_{i+1}-x_{{\mathrm{brink}},i}$ for the dunes in the measured field, we find that this quantity is between 1.5 and 4 times the height of dune $i$ at its brink, $H_{{\mathrm{brink}},i}$ (see inset of Fig. \ref{fig:interdune_distances}). These interdune distances can not be reproduced with the original transverse dune model proposed in \cite{Schwaemmle_and_Herrmann_2004}, where the separation bubble was defined in such a way as to have lengths between $6$ and $10$ times the dune height. Here we have to remark that the separation bubble defined in our model is the region with zero net flux at the lee side of the upwind dune. As mentioned above, the actual wind behaviour and, thus, the flux at the lee side have been found to present complex patterns before the reattachment point \citep{Walker_and_Nickling_2002}, a feature that is not captured in our model. On the other hand, as we can see in Fig. \ref{fig:interdune_distances}, interdune spacing also depends strongly on the position of the brink relative to the crest. Here, we plot the horizontal distances between the dunes, ${\Delta}_{i,i+1}$, measured from the foot of the slip face of dune $i$ to the beginning of the windward side of dune $i+1$, as a function of the distance between the brink of dune $i$ and the maximum of the parabola in Fig. \ref{fig:parabola}. In other words, dunes that have the position of the brink at the left (right) of the maximum of the parabola (point $C$) in Fig. \ref{fig:parabola} are associated with a proportionally larger (shorter) interdune spacing relative to the corresponding downwind dune. 

\cite{Parsons_et_al_2004} have found lengths of 3 to 15$h$ for the separation zone using CFD simulations, where $h$ is the height of the transverse dune. However, in this work, they only studied sharp crested transverse dunes, where the crest coincided with the brink. Simulations and measurements of surface shear stress over transverse dunes in a wind tunnel \citep{Walker_and_Nickling_2003} have reported similar reattachment lengths for such sharp crested dunes. In a recent work, the profile of the wind over the transverse dunes studied here has been calculated using FLUENT \citep{Herrmann_et_al_2005}. It was shown in this work that there is in fact a strong dependence of the length of the separation bubble with the shape of the dune at its crest, where dunes with a large crest-brink distance, i.e. very round dunes, almost do not show recirculating flow at its lee side. Moreover, \cite{McKenna_Neuman_et_al_2000} have noticed differences in the behaviour of the streamlines, flow expansion and deposition for sharp crested dunes in relation to the rounded ones.

The observations above lead to the conclusion that simply considering the length of the streamlines as proportional to the height at the brink is not reasonable for modelling transverse dunes, and that the observed dependence of the interdune spacing on the crest-brink distance of the dunes has to be taken into account. Since in our model the net sand flux is set to zero in the separation bubble, the observed interdune distances define an upper bound for the reattachment length. If we choose a larger length for the bubble in our model, simulation of closely spaced transverse dunes as the ones measured here is not possible \citep{Schwaemmle_and_Herrmann_2004}. The linear relation between the interdune spacing and the distance ${\Delta}_{\mathrm{brink-crest}}$ between the crest and the brink, which is shown in Fig. \ref{fig:interdune_distances}, suggests that, if the horizontal distance between brink and crest is described by a parabola of the form $H_{\mathrm{crest}}-H_{\mathrm{brink}} = k_1{{\Delta}_{\mathrm{brink-crest}}^2}$, then in a first approximation we could define the distance from the crest to the reattachment point to follow also a quadratic relation $H_{\mathrm{crest}} = k_2{(x_{\mathrm{reat}}-x_{\mathrm{crest}})}^2$. This relation is then used to determine the point of reattachment $x_{\mathrm{reat}}$ of the separation bubble, $x_{\mathrm{reat}} = {\sqrt{{H_{\mathrm{crest}}}/{k_2}}} + x_{\mathrm{crest}}$. In order to give an insight into the meaning of the parameter $k_2$, we notice that for a dune with height $7$ m at its crest, a value of $k_2 = 0.02$ gives $x_{\mathrm{reat}}-x_{\mathrm{crest}}= 18$ m, while $k_2 = 0.01$ leads to $x_{\mathrm{reat}}-x_{\mathrm{crest}} = 26$ m. 

In our previous work, the streamlines of the separation bubble were fitted by a third order polynomial, $f(x)=a_3x^3+a_2x^2+a_1x+a_0$, whose parameters were calculated from the continuity conditions of the profile $h(x)$ at the brink of the dune and at the reattachment point, as well as of $dh/dx$ at the brink. Moreover, the third order polynomial was calculated in such a manner that the separation bubble doesn't descend steeper than $14^{\circ}$ \citep{Schwaemmle_and_Herrmann_2004}. This leads to large values of the horizontal distance from the brink up to the reattachment point, $x_{\mathrm{reat}}-x_{\mathrm{brink}}$ being typically between 6 and 10 times the dune height. Using higher order polynomials to take into account the relation observed in Fig. \ref{fig:interdune_distances}, on the other hand, makes the streamlines descend too fast close to the brink at the lee side, making the curvature of the separation bubble different from that of the dune surface at the brink, thus giving rise to inconsistencies in the calculation of the shear stress with Eq. (\ref{eq:shear_stress_perturbation}) for smooth hills \citep{Weng_et_al_1991}, and consequently of the sand flux. 

\begin{figure}
\begin{center}
\includegraphics*[width=0.75\columnwidth]{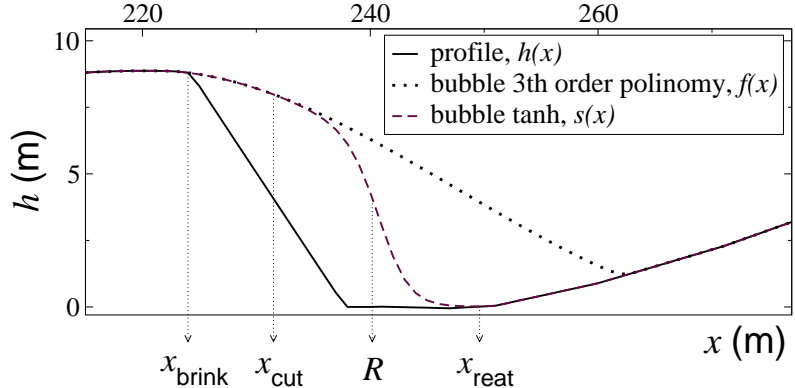}
\caption{Profile $h(x)$ of measured dune 3 with the modified separation bubble $s(x)$. The 3rd order polynomial separation bubble $f(x)$ is shown for comparison.}
\label{fig:sepbub}
\end{center}
\end{figure}

Therefore we define a modified separation bubble as  $s(x)=f(x)$ for $x_{\mathrm{brink}}<x<x_{\mathrm{cut}}$ and 
\begin{equation}
s(x)=\left[f(x)-h(x_{\mathrm{reat}})\right] \times {\left\{1-\tanh{\left[(x-R)/{{\lambda}_{\mathrm{c}}}\right]}\right\}}/2 + h({x_{\mathrm{reat}}}), \label{eq:sepbub}
\end{equation}
for $x_{\mathrm{cut}} \leq x < x_{\mathrm{reat}}$, where $x_{\mathrm{cut}}=x_{\mathrm{brink}}+{\zeta}({x_{\mathrm{reat}}-x_{\mathrm{brink}}})$, $R = (x_{\mathrm{cut}}+x_{\mathrm{reat}})/2$, ${\lambda}_{\mathrm{c}}$ determines the slope of the streamlines at point $R$, and $\zeta$ is a parameter between 0 and 1, which determines the point at which the original polynomial is multiplied by the function $\tanh$. The lengths involved in the modified separation bubble are schematically shown in Fig. \ref{fig:sepbub}. 

Eq. (\ref{eq:sepbub}) gives a shorter separation bubble from the {\em{original}} function $f(x)$. It keeps the slope of the original separation bubble close to the brink, and presents a steeper descent to the reattachment point. Thus, Eq. (\ref{eq:sepbub}) is a simple approximation of the original separation bubble introduced in \cite{Sauermann_et_al_2001} to study closely spaced transverse dunes, based on the phenomenological observations reported in the present work. We set $k_2=0.014$, which gives a distance of 25 m from the crest of dune 3 up to the reattachment point at its lee side, and ${\lambda}_{\mathrm{c}}$ and $\zeta$ are set to $0.1$ and $0.9$ respectively. 

Studies of the separation bubble of {\em{isolated}} transverse dunes have shown that the separation streamlines may be well described by an elliptical curve, but still separation distances only larger than 4 times the dune height were found, and no mention has been made to closely spaced transverse dunes in a field. Besides this, we report in the present work larger crest-brink distances than the ones of the dunes studied in \cite{Schatz_and_Herrmann_2005}. Simulations using our model with the separation bubble lengths calculated in \cite{Schatz_and_Herrmann_2005} would yield much more widely spaced transverse dunes than the ones investigated here. An extension of this work for transverse dune fields with dunes of different heights and crest-brink distances could yield more precise knowledge of the separation streamlines of closely spaced dunes. 

In the next section, we present simulations of transverse dune fields in two dimensions and compare the results with the measured profile in Fig. \ref{fig:profile}.

% **** SECTION 5: SIMULATIONS OF TRANSVERSE DUNE FIELDS IN TWO DIMENSIONS ****

\section{Simulations of transverse dune fields in two dimensions}

The numerical model described above has been used in \cite{Schwaemmle_and_Herrmann_2004} to study transverse dune formation on a sand bed completely filled with sand. As mentioned before, such dune fields developed on sand sheets have been classified as {\em{transgressive}} dune fields \citep{Hesp_et_al_1989}. For periodic boundary conditions, we choose at $t=0$ a sand bed of depth 10 m randomly disturbed by small Gaussian hills. We use the separation bubble introduced in the last section, which is able to give shorter distances between dunes than in the case \cite{Schwaemmle_and_Herrmann_2004}. 

\begin{figure}
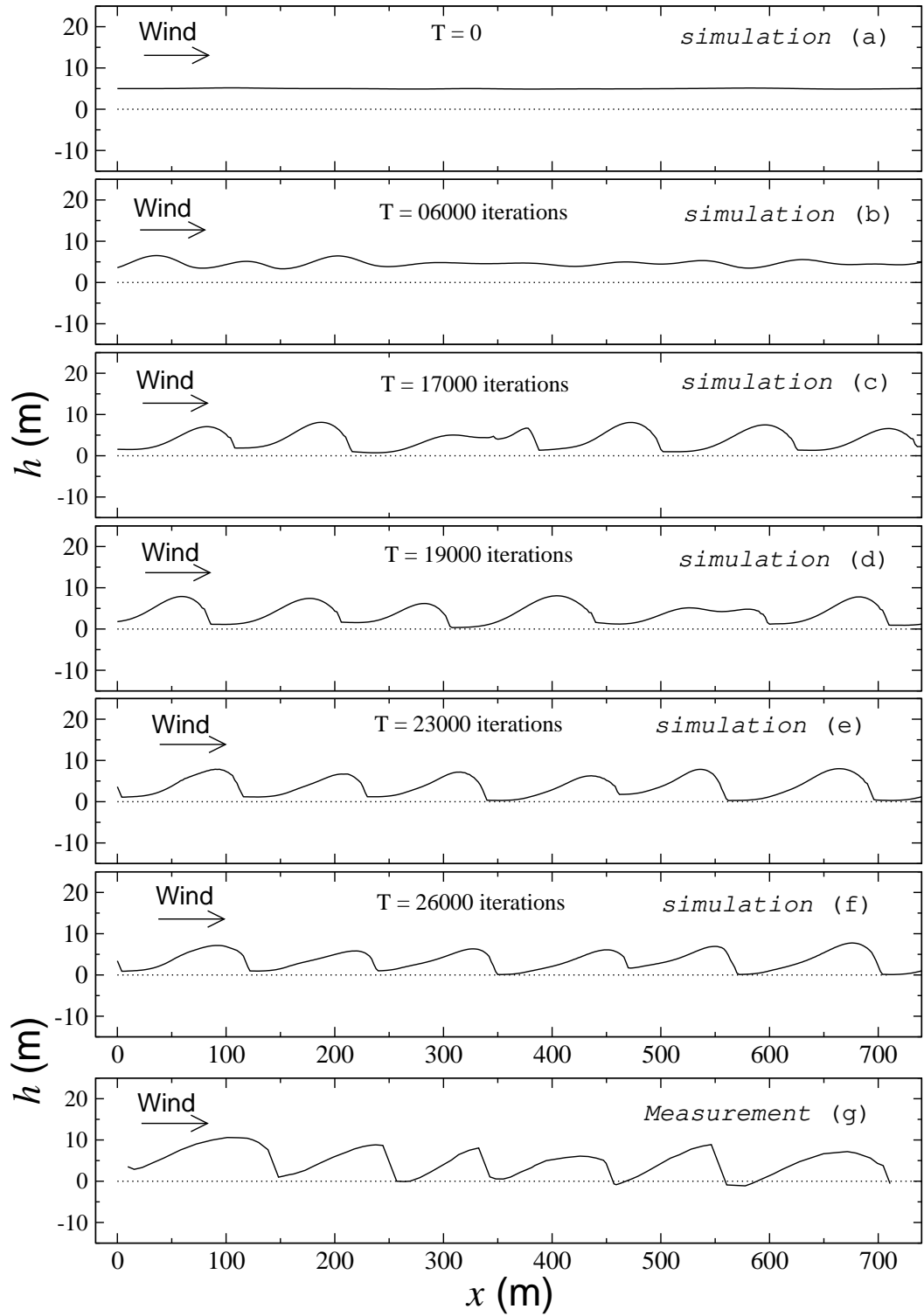

\begin{center}
\includegraphics*[width=1.0\columnwidth]{fig13a.eps}
\includegraphics*[width=1.0\columnwidth]{fig13b.eps}
\caption{From (a) to (f) we show snapshots of the simulation of a transverse dune field obtained from our two-dimensional model. Model parameters are mentioned in Section 4. The shear velocity $u_{{\ast}0}$ and its threshold $u_{{\ast}{\mathrm{t}}}$ are respectively $0.44$m/s and $0.32$m/s. For comparison, the measured profile is shown in (g).}
\label{fig:simulation}
\end{center}
\end{figure}

Figs. \ref{fig:simulation}(a) to (f) show simulation snapshots of a typical transverse dune field of 2 km length obtained with our two-dimensional model. The model parameters are those defined in Section 5. The values of the shear velocity $u_{{\ast}0}$ and its threshold value $u_{{\ast}{\mathrm{t}}}$ are respectively $0.44$ and $0.32$ m/s. For comparison, we show in Fig. \ref{fig:simulation}(g) the measured profile. The mean slope of the windward side in Fig. \ref{fig:simulation}(f) is between $0.065$ and $0.09$, which are values close to the windward side inclination found in the measured field ($0.08-0.12$). 

Furthermore, we see in Fig. \ref{fig:simulation} that the distance between crest and brink for the simulated transverse dunes of similar heights varies strongly as in the case of the measured dunes. This striking observation appears to be a consequence of the interaction between transverse dunes in a field. Barchans behave more like isolated dunes and exchange less sand than a field of closely spaced transverse dunes. Simulations in \cite{Schwaemmle_and_Herrmann_2004} have shown that the brink position of transverse dunes is in general closer to the crest if the dune is downwind of another one. Such sharp crested dunes with ``triangular'' cross-section are found in our simulations during the evolution of the field, as shown in Fig. \ref{fig:simulation}(c). However, they appear only when interacting with another dune like solitary waves, a feature that can be ilustrated comparing Figs. \ref{fig:simulation}(c) and (d). In order to investigate dune interaction in more detail, transverse dunes have to be still systematically studied as done recently for barchan dunes \citep{Schwaemmle_and_Herrmann_2003}. 

The typical dune shape found in our simulations is the ``rounded'' one, i.e. where the brink is clearly distinguishable from the crest. \cite{Tsoar_1985} has shown that the `rounded' dunes are `{\em{more stable}}' than the `triangular' ones, since the sharp-crested shape experiences strong flux at the brink. We have found such peaks of sand flux when trying to simulate the evolution of triangular transverse dunes with our model, and we have seen that they lose this shape after some iterations, becoming `round'. On the other hand, we could not find dunes with the `triangular' shape in our simulations of dune fields, but those interacting like solitary waves, as mentioned above. However, we do not conclude from this that the measured sharp crested dunes found in Fig. \ref{fig:profile} should be interacting like solitary waves with their neighbours. We remark that our model assumes lateral invariance, from which the profile variations found for these dunes in the $y$ direction, as we see in Fig. \ref{fig:field}, are not included and, consequently, variations in the lateral shear stress are not considered, which could influence the dune shape at the brink. In spite of this limitation, the variable crest-brink distances found in the field have been captured by the model. Also, the separation bubble in our model is the region of zero flux downwind, which implies that the observed interdune distances provide an upper bound of the separation length. To use a longer bubble, our model could be improved in order to capture the complexity of the wind and flow in the bubble \citep{Frank_and_Kocurek_1996,Wiggs_2001,Walker_and_Nickling_2002}. Possibly, in this case, sharp crested dunes could be found in simulations of closely spaced transverse dunes.

The shape of coastal dunes like the ones investigated here may be influenced by standing water and vegetation in the interdune area, which could be relevant to the lee face steepness and crest-brink separation distances. Recently, an extension of our model to simulate barchan dunes with vegetation has successfully reproduced the observation that these dunes transform into parabolic dunes \citep{Duran_et_al_2005}, as suggested in \cite{Tsoar_and_Blumberg_2002}. The study of coastal transverse dune field formation in the presence of vegetation and lagoons is the subject of further work.

% *******************  SECTION 6: CONCLUSIONS ********************************

\section{Conclusions}

We have performed field measurements of the height profile of transverse dunes in the Len\c{c}\'ois Maranhenses, northeastern Brazil, and compared the results with the predictions of a two-dimensional dune model recently introduced. The model separation bubble at the lee side of the dune was adapted in order to match the observation that the interdune spacing in the measured field is less than 4 times the dune height at the brink.

The shape of transverse dunes was found to vary with their height in a different way than in the case of barchan dunes. While larger barchans present their crest closer to their brink than smaller ones, the crest-brink distance was found to be independent of the height for transverse dunes. Moreover, we found that the windward side of all dunes could be superposed in such a way to form a nearly parabolic shape with common maximum point (see Fig. \ref{fig:parabola}), independently of the relative position of the brink for each particular dune. Based on the relation observed for the interdune distances as a function of the brink position, we defined the reattachment length at the dune lee side to simulate transverse dune fields similar to the measured one. Our model successfully reproduces the aspect ratio of the measured dunes, interdune distances and the appearence of variable crest-brink distance, assuming lateral invariance of the profile, which allows considerable simplifications of the model and simulations in two dimensions. In order to consider the oscillations in the profile in the $y$ direction, as Fig. \ref{fig:field}(b) shows, three dimensional simulations are required to take into account the lateral shear stress \citep{Schwaemmle_and_Herrmann_2005}, which could of course influence the dune shape for instance at the brink. The simulated dunes compared with the measured ones appeared `rounded', and we have discussed possible explanations for this.

Finally, a small field in the beginning of a growing structure of transverse dunes was measured which characterizes the pattern of the Len\c{c}\'ois Maranhenses, as illustrated in Fig. \ref{fig:field}(a). We have simulated a transverse dune field that evolved on a sand sheet, thus a ``transgressive'' dune field \citep{Hesp_et_al_1989}. One aspect our model did not take into account is the proximity of the water line to the surface due to the rainfall in this region and the eventual role of standing water or vegetation, which may exert a major influence on the dune shape, although in the dry seasons water is practically absent from the sand sheet where the measured transverse dunes were found. Comparison of our field data with further field measurements on transverse dunes in the Len\c{c}\'ois Maranhenses as well as in other areas with different climate and geological conditions could, thus, be of importance for better understanding formation and evolution of transverse dune fields.

\begin{ack}
We acknowledge O. Dur\'an and A. O. Sousa for discussions and V. Schatz for his interest in this work and helpful comments. We also acknowledge H. Besler for helpful comments. We acknowledge Juliana Cristina Fukuda of the administration of the National Park of the Len\c{c}\'ois Maranhenses and IBAMA/Brazil for useful informations about the Len\c{c}\'ois Maranhenses. This work was supported in part by Volkswagenstiftung and the Max-Planck Prize awarded to H. J. Herrmann (2002). E. J. R. Parteli acknowledges support from CAPES - Bras\'{\i}lia/Brazil.
\end{ack}

\newpage

% ***********************************************************************
% **********************        REFERENCES   ****************************
% ***********************************************************************

\end{document}